\documentstyle[preprint,aps,epsf]{revtex}
\begin{document}

\newcommand{\be}{\begin{equation}}
\newcommand{\ee}{\end{equation}}
\newcommand{\BE}{\begin{eqnarray}}
\newcommand{\EE}{\end{eqnarray}}
\newcommand{\BEn}{\begin{eqnarray*}}
\newcommand{\EEn}{\end{eqnarray*}}
\newcommand{\barr}{\begin{array}} 
\newcommand{\earr}{\end{array}}
\newcommand{\bit}{\begin{itemize}}      
\newcommand{\eit}{\end{itemize}}
\newcommand{\bfl}{\begin{flusleft}}
\newcommand{\efl}{\end{flusleft}}
\newcommand{\bfr}{\begin{flushright}}
\newcommand{\efr}{\end{flushright}}

\newcommand{\bc}{\begin{center}}
\newcommand{\ec}{\end{center}}

\newcommand{\ben}{\begin{enumerate}}    
\newcommand{\een}{\end{enumerate}}

\newcommand{\cl}{\centerline}
\newcommand{\ul}{\underline}
\newcommand{\nl}{\newline}


\newcommand{\impl}{\Longrightarrow}
\newcommand{\eps}{\varepsilon}
\newcommand{\de}{\partial}


\newcommand{\xb}{{\bf x}}
\newcommand{\rb}{{\bf R}}
\newcommand{\kb}{{\bf k}}
\newcommand{\qb}{{\bf q}}
\newcommand{\eb}{{\bf E}}

\tightenlines
\draft

\pagestyle{plain}
\sloppy
\baselineskip 0.5cm

\title{FUZZY PHASE TRANSITION  IN A 1D COUPLED STABLE-MAP LATTICE}

\author{F. Cecconi$^{1,2}$, R. Livi$^{1,3}$ and A. Politi$^{2,4}$}
\vspace{1.cm}

\address{$(1)$ Dipartimento di Fisica, Universit\`a di Firenze, 
               L.go E. Fermi 2, 50125 Firenze, Italy\\   
         $(2)$ INFN Sezione di Firenze,
               L.go E. Fermi 2, 50125 Firenze, Italy\\   
         $(3)$ INFM Unit\`a di Firenze
               L.go E. Fermi 2, 50125 Firenze, Italy\\   
         $(4)$ Istituto Nazionale di Ottica, Firenze, 
               L.go E. Fermi 6, 50125 Firenze, Italy\\}   
\date{\today}
\maketitle

\begin{abstract}
A coupled-map lattice showing complex behaviour in presence of a fully 
negative Lyapunov spectrum is considered. A phase transition from
ordered to disordered evolution, upon changing
diffusive coupling is studied in detail. Various
indicators provide a coherent description of the scenario: the existence of
an intermediate transition region characterized by an irregular alternancy 
of periodic and chaotic evolution. 
\end{abstract}

\vskip 1.cm

\section{Introduction}

Very often the approach to space-time chaos in spatially extended systems
is based on the extension of concepts and tools developed for finite (low)
dimensional systems. For instance, dynamical indicators such as the
Kaplan-Yorke dimension and the Kolomogorov-Sinai entropy \cite{RE} have been 
turned into the corresponding intensive indicators, i.e. dimension
and entropy {\it densities} \cite{Gr}. This strategy is essentially motivated 
by the hypothesis that the dynamics of chaotic extended systems can be viewed 
as that of many, almost independent, finite-dimensional subsystems. Although
the existence of a limit Lyapunov spectrum \cite{Kan}
provides strong support to such
an idea, it is still rather unclear in which sense the evolution of
different pieces of, say, a chain of maps is truly uncorrelated.

Even more important is the observation that the infinite dimensionality of
the phase-space can give rise to entirely new features the understanding of
which requires different tools and perhaps will open new perspectives.
One such example that will be thoroughly studied in the present paper is the
occurrence of chaotic evolution in the presence of a negative maximum
Lyapunov exponent. This is indeed a phenomenon that can exist only in an
infinite dimensional phase space, as it can be shown with a simple
argument based on a {\it reductio ad absurdum}. An aperiodic evolution
requires that the limit set of a generic trajectory contains infinitely many
points. If the evolution is confined to a bounded region, there must be at
least one accumulation point. Now, since a sufficiently small box centred
around any accumulation point contracts in all directions (the maximum
Lyapunov exponent being negative), all trajectories in the vicinity of the
accumulation point are asymptotically indistinguishable and there can be at
most a periodic cycle. This argument breaks down if we have to consider
an infinite dimensional phase-space, since in this case the limit set can
well be made of infinitely many points, all within a bounded region and yet
a finite distance from one another. The above is not only a theoretical
possibility, but a feature actually observed in several models such
as coupled maps \cite{CK88,PLOK,B} and oscillators \cite{BP}, although no 
detailed explanation of the
underlying mechanisms has yet been provided. What is most striking of this
phenomenon is the empirical evidence that the evolution of large enough
systems appears to be irregular and stationary in time, so
that it makes sense to speak of a ``Lyapunov stable'' chaotic regime. In the
following, we shall use the short-hand notation {\it stable chaos} (SC) to
identify this type of behaviour.

In a previous paper it has been shown that SC is a robust phenomenon in
the sense that it persists in finite regions of the parameter space \cite{PLOK} 
and it survives even if the discontinuities in the dynamical equations are 
removed \cite{Russi,PT}. Also the discreteness of the time variable does not 
seem to be a severe limitation as SC has been observed also in a chain of 
coupled Duffing oscillators \cite{BP}. The only true limitation seems to be 
the need for a synchronous external forcing of the oscillators.

SC can be partly understood by unveiling the analogy with actual simulations
of chaotic maps on digital computers. Any computer has a finite accuracy
which is determined by the number of bits used in the internal
representation of a real number. As a consequence, even a chaotic map sooner
or later must yield a periodic orbit. This apparent limitation has not
prevented an effective use of digital computers in the study of
deterministic chaos. In fact, if the computer-word is sufficiently long,
the time required for observing the collapse onto a periodic cycle is so
long that this ``transient'' regime is almost indistinguishable from the
truly stationary regime of the chaotic map. If one substitutes the length of
the computer-word with the spatial length of an SC system, we can rephrase
the above arguments and thus provide indirect support for the existence of a
stationary chaotic regime in infinitely extended systems. However, it is
honest to recognize that in the case of deterministic chaos there exists a
well developed theory \cite{RE} which, starting from Smale horseshoe and 
Anosov systems, predicts the occurrence of irregular behaviour in mappings 
over the real numbers. In this case, one is faced only with the 
problem of explaining why an actual simulation reproduces almost exactly the
theoretical expectations. Conversely, for what concerns SC, there is mainly
numerical evidence and no theory stating that under some specific 
circumstances one can expect a chaotic evolution in an infinitely extended 
system. The only pieces of a theory can be constructed at the
expence of a further simplification which is, however, very enlightening.
If one discretizes the continuous state-variable in a chain of maps, it is
very natural to invoke an analogy with deterministic cellular automata
(DCA). In fact, DCA too can exhibit chaotic behaviour only in the
infinite-size limit, the finiteness of the number of possible states ($2^L$
for binary automata) necessarily implying the eventual convergence to some
periodic orbit. The correspondence between chaotic DCA and SC can be put on
a more rigorous ground by first encoding the patterns originated by an SC
regime and then trying to reproduce them by some DCA with a suitable range
of interaction. The first step is nothing but an implicit statement on the
existence of a generating partition. The work done in low-dimensional
systems has shown that rather coarse partitions can be constructed which
reproduce the dynamics of chaotic maps without loss of 
information \cite{GK,CP}. Thus,
we do not expect this step to be particularly harmful in the context of
SC, the only possible problem being the actual construction of a generating
partition in specific cases.

The second step is not obvious at all, since it is not known to what
extent a pattern with no local production of information can be reduced to a
DCA. Let us start the discussion of this issue by recalling that
low-dimensional chaotic systems, such as the H\`enon map, are equivalent to
probabilistic automata, where the probability of the next symbol effectively
depends on some previous symbols (their number corresponding to the order of
the Markov process). The probabilistic character of the automaton is
intrinsically related to the existence of an expanding direction and to
the corresponding amplification of uncertainty. The coupling of chaotic
maps, as it occurs in spatially extended systems, leads naturally to
probabilistic {\it cellular} automata: the probability of a symbol in a
given place at a given time does depend not only on the past symbols in the
same site as in the previous case, but also on the past symbols in the
neighbouring sites.

In the case of SC, there is no local amplification of uncertainty, so that
it is tempting to conjecture that the future symbol is exactly determined,
once the past history of all previous symbols is known. This hypothesis
has been already tested in several cases, finding that it would be more
appropriate to state that it is the whole new configuration to be predicted
with almost no uncertainty. However, in some cases, it has been found that a
DCA with a long enough space-time memory suffices to reproduce exactly
the observed pattern, while in other cases, the uncertainty of each
forecasted symbol decreases and presumably goes to zero only in the limit of
an infinite range of interaction. Therefore, it is definitely reasonable to
affirm that DCA represent a subclass of SC systems and what is known about
DCA can be automatically translated into the language of SC. Leaving aside
the question whether SC encompasses some type of behaviour absent
in DCA, here we want to stress the important advantage of SC over DCA: the
existence of a tunable continuous control parameter. Such a possibility
is particularly appealing in view of the conjectured existence of
``complex'' behaviour at the edge of chaos \cite{Lang}. In fact, it has been 
suggested that a true richness of behaviour is observed whenever the 
underlying rule of a DCA is in some sense halfway between ordered and 
chaotic rules. However, testing of the above idea requires a continuous 
parametrization of all the rules. Since DCA rules are intrinsically discrete, 
different artificial procedures have been introduced, and each of the 
proposals is weak in some respect. In the context of SC, the continuous 
parameters, naturally present in the original model, allow one to study 
the very same question without having to bother with the problem of 
introducing {\it ad hoc} a suitable parameter.

In fact, the question of how we pass from an ordered to a chaotic regime
in SC systems is perfectly legitimate, as revealed by simulations performed
for different choices of the control parameter. 
Although in the past much attention has been pointed 
to the occurrence of chaotic rather than to
the natural occurrence of periodic evolution, both regimes can arise 
and, in our opinion, it is very important to shed light on the transition
between these two regimes: is that a standard thermodynamic phase-transition,
or  do we find the signature of ``complexity''? Or is it even as simple as
a ``bifurcation''?

The order-to-chaos transition suggests also a comparison with standard
space-time intermittency (STI) occurring in chaotic systems\cite{ChMa}. The 
latter
phenomenon has been shown to be strictly related to directed percolation
transition. A posteriori, this is not very surprising since, on the one 
hand, a locally chaotic evolution is reminiscent of probabilistic cellular
automata (see above), while the ordered  dynamical configuration  
can play the role of an absorbing state. However, the analogy has been shown 
not to be a complete equivalence between the two phenomena, since 
finite regions characterized by
chaotic behaviour cannot be assimilated to truly stochastic domains. A
reminiscence of the alternancy of regular and irregular behaviours - typical
of low-dimensional systems - indeed survives, leading to a more ``exotic''
evolution on the ordered side of the phase transition \cite{GrSc}~. Now, an
order-to-chaos transition occurring in  an SC system should exhibit even more
striking deviations from a percolation transition, basically because any
finite ``chaotic domain'' cannot be chaotic at all! This is a first
indication that the transition cannot be a ``simple'' equilibrium
phase-transition, as the studies described in this paper will confirm.
However, the link with STI is more subtle than one could naively
think. It was already shown that STI can be effectively described by
a sequence of DCA constructed by suitable discretization of the local 
dynamics \cite{ChMa}.
In fact, any DCA can be seen as a stepwise map: the smaller is
the separation between consecutive steps, the more accurate is the
reproduction of the dynamics. Upon changing the control parameter, one
passes discontinuously from one to another rule. Thus, for any finite
resolution, in a finite number of steps (changes of rule), one passes from
ordered to chaotic behaviour: no truly continuous parametrization is
recovered unless the limit of infinitely many symbols is taken, i.e.
the continuous nature of the local variable is restored. However, given
any stepwise representation of the local dynamics, one could proceed in
a different manner, tilting each of the steps of the local function. In this
way, the continuous nature of the variable is immediately restored and if
the slopes of the various steps are not too large, the maximum Lyapunov
exponent is bounded to be negative. Qualitatively speaking, we have the same
phenomenon as in STI, a transition from ordered to chaotic behaviour.
Quantitatively, in this paper we conjecture that a ``complex''
region is expected to arise in parameter space of SC systems.

Finally, it is worth mentioning another similar transition, extensively
studied in the context of neural and Kauffman networks \cite{Kau69}. 
There, the state
variable is discrete (typically binary) and the evolution rule is
entirely deterministic exactly as for DCA. At variance with DCA, there are:
\begin{itemize}
\item{i)} Quenched disorder: the updating rule operating in a given cell
(synapsis) is randomly chosen.
\item{ii)} Lack of topology: each cell interacts (is connected) with a
randomly chosen set of other cells.
\end{itemize}
In such a context, it has been found that upon decreasing, e.g., the
correlation between synaptic couplings (a continuous parameter in
the thermodynamic limit), a transition occurs from a chaotic regime to
frozen patterns. The transition appears to be a ``standard'' continuous
order-to-chaos transition located at a specific value of the control
parameter. Within the paradigm of a meaningful complexity occurring at the
edge of chaos, it has been conjectured that the most meaningful choice of
the parameters for the network to be a realistic model of either gene 
regulation or neural activity is close to criticality\cite{Kau}. While we are 
not going to comment about such speculations, here we shall investigate the
nature of the order-to-chaos transitions occurring in a 1d lattice of stable
maps, finding evidence that the transition region is rather intricate and 
highly irregular.

A somehow similar phenomenon has been already investigated in a 2d lattice
of stable maps, finding correspondence with a nonequilibrium transition 
from weak to strong turbulence \cite{CLP}. In such a case, it has been 
possible to reproduce the key features of the entire phenomenon by means of a 
suitable stochastic equation \cite{KLP}. We suspect that such a transition is 
indeed close to a true stochastic process since, even in the most ordered 
(weak turbulence) regime,
there are infinitely long interfaces which, in spite of the local stability,
can be characterized by a pseudo-random evolution. In fact, it is the infinite 
dimensionality of the phase-space that makes possible the generation of an 
irregular behaviour over an infinite time lag.

This paper is arranged as follows. In section II we present the model,
recalling the features of SC and giving a brief overview of the
phenomenology occurring for various values of the coupling strength
(our control parameter). In section III we study  
 space-time correlation functions and perform 
damage-spreading analysis since they both allow to identify a proper order 
parameter for the transition. In section IV, information-theoretic 
concepts are introduced for a supplementary investigation. 
Section V is finally devoted to discussions and conclusions.

\section{A model of stable chaos} 

The dynamical system considered in this paper is a one-dimensional lattice 
of diffusively coupled maps
\be
x_i(t+1) = (1-2\eps) f(x_i(t)) + \eps \big[ f(x_{i-1}(t)) + 
f(x_{i+1}(t))\big]  \quad ,
\label{eq:cml}
\ee
where $\eps \in [0,1/2]$ is the coupling constant and periodic
boundary conditions are assumed over a length $L$. The local mapping has 
the form
\be
f(x) = \left\{ \barr{lc}
	     bx \quad ,& \;\; 0 < x < 1/b       \\
                &                        \\
  a + c(x-1/b)  \quad ,& \;\; 1/b < x < 1
\earr  \right.
\label{eq:map}
\ee
One can easily realize that this mapping can yield stable periodic
dynamics for sufficiently small values of $c$. In what follows,
we fix the set of parameter values
~$\{a=0.07,\,b=2.70,\,c=0.10\}$ in such a way that,
for any initial condition $x \neq 0$, the attractor of the
local mapping is a stable period-3 orbit.

It is worth stressing that the stability of local dynamics (\ref{eq:map}) 
implies the stability of map (\ref{eq:cml}), whose maximum Lyapunov exponent 
turns out to be negative for any value of $\eps$. 
As a consequence, the long-time evolution of the diffusively coupled
system is confined to a periodic attractor. 
Despite  this constraint, we are going to show that very different 
dynamical regimes can be observed, depending on the coupling $\eps$.

A space-time representation of the evolution can be obtained by 
encoding the variable $x$ with suitable gray-levels. Some typical patterns of 
the different regimes are reported in Fig.~\ref{fig:patt}. An ordered
pattern is basically a random arrangement of a few different ``stripes'', each 
stripe corresponding to a periodic orbit (see Fig.~1a,c). While the 
trajectories are always
stable against infinitesimal perturbations, the global stability properties 
are governed by the behaviour of the domain walls (DWs) separating the various 
stripes (either different phases of the same solutions or truly
different orbits).

The key properties of the evolution appear to be related precisely to the  
DW dynamics. Depending on the parameter values, it may happen that the DWs
move with different velocities and interact with each other according to 
different collision rules. Upon changing $\eps$, these rules change as well as 
the structure and the number of DWs. If, at the end, only a population of DWs
survives, all moving with the same velocity (in particular, 
stationary ones), then we still have an ordered evolution\cite{comm}. 
Alternatively, one has a ``gas'' 
of DWs moving with different velocities, i.e. a sort of chaotic evolution: the 
DWs may even be so dense that it is impossible to properly identify them
(see Fig.~1d). Moreover, there are cases in which the asymptotic 
(time) periodic pattern is not attained over the available integration laps.

Even the reader vaguely acquainted with the dynamics of DCA should have 
recognized in the above sketched regimes the various classes of such  models
\cite{Wol}. Therefore, it is definitely tempting to use model (\ref{eq:cml}) 
for checking the existence of a complex phase separating ordered from chaotic 
motion.

As it has been already discussed in Refs.~\cite{buni,PLOK} a criterion for
distinguishing chaotic from ordered behaviour is provided by the scaling
properties of the transient duration with the chain length, starting from
random initial conditions. Notice that this approach is the same adopted in 
the characterization of the order-to-chaos transition occurring, for instance,
in neural networks \cite{BaPa}. The transient duration is defined as the 
number of iterations necessary to observe the first {\it recurrence},
\be
  T_r(L) = \hbox{min} \big \{t \,\,|\,\, d(\{x\}_t,\{x\}_\tau) < \delta, \tau<t
	\big \} \quad ,
\ee
where $d(\{x\}_t,\{x\}_\tau)$ is the distance 
between the configurations at time $t$ and $\tau$, respectively, computed 
using some specific norm (here we considered the maximum norm). 
All the conclusions 
hereafter reported are independent of the actual value of the parameter 
$\delta$, provided it is small enough ($\delta$ has been fixed to 
$10^{-4}$ in all our simulations). As it was already noticed in 
Ref.~\cite{PLOK}, the chaotic regime can be identified by the exponential 
growth with $L$ of $\langle T_r(L) \rangle $, where the average
$\langle \cdot \rangle$ is performed over the ensemble of random initial 
conditions.

A global picture of the the average transient time $\langle T_r(L) \rangle$ 
is shown in Fig.~2a for different values of $\eps$. 
The strong variations in the order of 
magnitude of $\langle T_r(L) \rangle$ do confirm the visual impression of an 
irregular alternancy of ordered and chaotic regimes. This is further 
strenghtened by the comparison between the solid and the dashed curves 
(corresponding to $L=50$ and $40$, respectively), that single out the chaotic 
regions as those where the solid curve is consistently above the dashed one.

Before discussing the various approaches used for investigating the transition 
region, let us comment about another aspect of the evolution of finite chains:
the period $T_p(L)$ of the asymptotic state. In principle, $T_p(L) < T_r(L)$;
in practice, $T_p(L)$ can be much shorter, as seen in Fig.~2b, where the
average period $\langle T_p(L) \rangle $ is reported versus $\eps$, showing
strong fluctuations, while it may remain  rather ``short''  
deeply inside the chaotic regions. It is worth stressing that this 
phenomenology is completely different from what observed in neural networks, 
where the chaotic phase is characterized also by periods as long as
transients \cite{BaPa}. Nonetheless, in model (\ref{eq:cml}), one observes
an accumulation of longer and longer periods when any transition
is approached from the ``ordered'' side.
We shall comment more carefully on this point in 
the next section.

\section{Characterization of the phase transition}

Direct inspection of  Figs.~2  shows that the widest
chaotic region is approximately located in the interval $\eps \in
[0.3\, , \, 0.4]$. Incidentally, it is in this region that the
first evidence of SC was found in this model for $\eps = 1/3$
(see Ref.\cite{PLOK})~. For this reason we have chosen to point our
attention to the parameter region close to $\eps = 0.3$.
Transient analysis and spatio-temporal patterns obtained from the simulations 
give a clear evidence that, sufficently below $\eps = 0.3$, there exists a 
periodic phase (PP), whose dynamical properties seem to change continuosly with 
the control parameter. Analogously, sufficiently above $\eps = 0.3$ 
there exists a structurally stable chaotic phase (CP). 
In other words we are in presence of a phase transition of order/disorder type.
In what follows we shall characterize it by analyzing the
behaviour of some observables, aiming also to indentify an order
parameter.
More precisely, in the first subsection, we discuss the properties of 
the spatio-temporal correlation functions, finding that only the  CP
displays a temporal decay to zero. 
In the second subsection, we study the propagation of initially localized 
perturbations (damage-spreading analysis), that is found to drop
to zero in the PP.
A careful application of the above tools has consistently revealed that
there is not a single threshold separating the two phases,  but 
rather a whole ``fuzzy'' region $\eps \in [0.3,0.3005]$, where periodic and
chaotic behaviours alternate in an apparently irregular manner.

We believe that the peculiarity of this transition should be attributed
not only to the deterministic nature of the model (as in the
case of STI), but also to the specific absence of a local source of chaos. 

\subsection{Correlation Functions}

Space-time correlation functions are common tools for describing the 
statistical properties of the motion in spatially extended systems.
In fact, they provide a first quantitative criterion apt to classify the 
various regimes observed in the dynamics of model (\ref{eq:cml}). 
In particular, they allow one to check whether the phase transition 
can be associated with the appearance of spatial long-range order.
 
The spatio-temporal correlation function is defined as 
\be
C(i,j;t,\tau) = \langle x_i(t)
x_{i+j}(t+\tau) \rangle_{t} -
\langle x_i(t) \rangle^2 \quad ,
\label{eq:corr}
\ee
where the average $\langle \cdot \rangle$ is performed over  
initial conditions made of independent, identically and uniformly 
distributed random variables $x_i(0)$. 

In view of the periodic boundary conditions and  because of the 
translational invariance of the initial conditions, $C(i,j;t,\tau)$ does 
not depend on $i$.
Conversely, for what concerns the dependence on $t$, there is no reason 
a priori for it to be irrelevant. On the other hand, if
the system approaches a stationary regime for sufficiently large $t$,
this dependence is practically negligible.  
The only case where this does not occur is PP, when the phase of the 
time periodicity still plays a role. 
As in the present investigation such a dependence is not relevant,
from here on we drop the dependence on $t$ in (\ref{eq:corr})~.

Moreover, since numerical simulations show the absence of travelling 
structures in the asymptotic configurations, here we can consider 
separately the spatial and temporal behaviour of (\ref{eq:corr}). 

Upon these remarks, we define the spatial correlation 
function as 
\be
C_S(j) = \langle x_i(t)
x_{i+j}(t) \rangle_{t} -
\langle x_i(t) \rangle_{t}^2  \quad ,
\label{eq:corrx}
\ee
where the subscript $t$ indicates that the average is performed also 
over time.
Independently of $\eps$, $C_S(j)$ exhibits an exponential decay over a 
few lattice units (see Fig.~\ref{fig:corrx}), indicating that, presumably,
spatial long-range order does not occur at the transition. 
This is compatible with the qualitative picture suggested by the patterns, 
which all present spatial disorder.
This behaviour makes doubtful the perspective of considering the present 
phase transition as a non-equilibrium critical phenomenon. 

The evolution of model (\ref{eq:cml})  can be characterized 
by the behaviour of the  
temporal correlation function
\be
   C_T(\tau)  = 
\langle x_i(t) x_i(t+\tau) \rangle_{t} -
\langle x_i(t) \rangle_{t}^2  \quad .
\label{eq:corrt}
\ee

In order to improve the statistics, in the numerical simulations
we have also performed an average 
over lattice sites separated by a distance larger than the spatial correlation
length. 

At variance with $C_S(j)$, $C_T(\tau)$ shows a very sensitive dependence 
on $\eps$, especially when approaching the critical zone, thus confirming 
the existence of different phases in the evolution of the system. 
The scenario can be summarized as follows.
In the whole range of $\eps$-values that we have considered, $C_T(\tau)$ 
displays period-two oscillations that appear to originate from the
abundancy of stable period-two orbits.
For $\eps$ below the lower ``threshold'' $\eps_c(1) = 0.3$, $C_T(\tau)$ 
does not decay (see Fig.~\ref{fig:corrt}a). This confirms the presence of a 
well defined phase, corresponding to a time periodic but spatially disordered 
dynamics (PP).

For values of $\eps$ inside the range $[\eps_c(1),\eps_c(2)]$, $C_T(\tau)$ 
may either decay to zero, as in CP (see below), or towards a non-zero 
asymptotic value (Fig.~\ref{fig:corrt}b).
To our knowledge, the only example in dynamical systems
of the latter behaviour occurs in period-two windows of chaotic maps,
where the phase point oscillates coherently between two different regions,
but irregularly inside them \cite{Ott}.
This is tantamount to interpreting the dynamical process as
small-amplitude fluctuations on top of a period-two signal. 
This interpretation seems to be quite reliable, as we have verified that,
in this region of $\eps$-values, all periodic orbits have an even period and
are characterized by a prominent period-two component. The crucial point is 
to understand whether the observed dynamics is exhausted by the superposition
of a finite number of different periodic components even in the limit of
$L \to \infty$. Only if this were not the case, the above interpretation 
could hold true. A convincing answer to this question can be given only
by carefully looking at the structure of the corresponding space-time patterns
which are anyhow made of periodic stripes, as in PP. On the other hand, the
observed growth of the average period $\langle T_p(L) \rangle$ (see Sect. 2)
does not allow for settling this controversial point on a purely numerical
ground. This point will be reconsidered in the next section.

For $\eps \geq \eps_c(2) = 0.3005$, $C_T(\tau)$  decays to zero (see, 
for instance, Fig.~\ref{fig:corrt}d). A good observable for the 
characterization of the transition from  CP to PP
is represented by the exponential decay-rate $\theta$ of the envelope of 
the temporal correlation function,
\be
{\rm Env}\{ C_T(\tau)\} \sim \exp(-\tau/\theta) \quad .
\ee
In fact, $\theta$ becomes very large as soon as $\eps$ approaches 
$\eps_c(2)$ from above. In this perspective, $\theta$ plays the role of an
order parameter, describing at least one side of this peculiar 
phase transition, where a ``fuzzy'' region separates the two phases.

\subsection{Damage spreading analysis}

The irregular dynamics observed in our model is produced by transport rather 
than by local production of information, which is not present in view of a
negative maximum Lyapunov exponent. The similarity with DCA, discussed in
the Introduction, suggests that the main features of this transport 
mechanism can be analyzed by studying the propagation of finite 
disturbances. In DCA language, this is the so-called damage-spreading 
analysis. In practice, it amounts to determine the effects produced on the 
pattern by localized perturbations of the initial state. An unbounded growth
of the region affected by the perturbation is usually considered as an
indication of a chaotic evolution \cite{PB}. 
In fact, this means that disturbances
arising at the boundaries can travel undamped through the whole system.

A perturbation can be introduced in the following way. Let
~${\cal X}_1 = \{x_1,x_2,....,x_L\}$~ and
~${\cal X}_2 = \{y_1,y_2,....,y_L\}$~ represent two initial configurations
such that
$$
y_i = \left\{\barr{lc}
        x_i + \delta_i & \;\; |i-L/2| \leq S   \\
                       &                       \\
        x_i            & \;\; \mbox{elsewhere}       
       \earr  
\right.
$$
where $\delta_i \sim {\cal O}(1)$, such that nonlinearities can play an 
effective role (the only nonlinearity present in our model is
the discontinuity in the map); $L$ is the chain length and $S$ is the size 
of the region where the two configurations are initially different. 
Then, ${\cal X}_2$ is said to be a perturbation of ${\cal X}_1$.
Typical damage-spreading patterns close to and inside the transition region 
are shown in Fig.~\ref{fig:def-patt}. Direct inspection indicates that 
an effective spreading occurs in CP (see Fig.5d)~. 

The transmission rate of information is then measured as the average velocity 
of increase of the perturbation size. In practice this can be 
defined by making reference to two different quantities:
\begin{itemize}
\item[(a)] the position of the perturbation front;
\item[(b)] the distance between two configurations.
\end{itemize}
In case (a) one first defines the left and the right fronts of the 
perturbation
\BE
F_l(t) & = & 
\min \bigg\{ 1 \leq i \leq L :\; |x_i(t) - y_i(t)| > 0 \bigg\} \nonumber \\ 
F_r(t) & = &
\max \bigg\{ 1 \leq i \leq L :\; |x_i(t) - y_i(t)| > 0 \bigg\} \quad.
\label{front}
\EE
In so far as the left-right spatial symmetry is not broken, as in the present
model, the two definitions are equivalent. The corresponding front
velocity is therefore
\be
V_F 
= \lim_{t\to\infty} \frac{\langle F_{l,r}(t)\rangle}{t} \quad.
\label{eq:velf}
\ee
In  case (b) the distance $D$ between two configurations of 
the system is given by
\be
D(t) = \frac{1}{L}\sum_{i=1}^{L} |x_i(t) - y_i(t)|
\ee
(notice that $D$ is a straightforward generalization of the Hamming distance
usually adopted in the study of DCA behaviour). 
Accordingly, the corresponding average damage-spreading velocity reads 
\be
V_D 
= \lim_{t\to\infty} \frac{\langle D(t) \rangle}{t} \quad.
\label{eq:veld}
\ee
In both the above definitions, $\langle \cdot \rangle$ denotes
the average performed over initial conditions. Positive values of $V_{F}$ and 
$V_{D}$ indicate that any two nearby configurations tend to separate in time.
In this sense, damage-spreading analysis can be considered
analogous to the Lyapunov stability analysis \cite{PB}.
 
In numerical simulations we have averaged over ~$500$~ initial conditions in 
order to reduce fluctuations. The results reported in Fig.~\ref{fig:vel} show 
that $V_F$ and $V_D$ are equivalent modulo a scale factor. 
This implies that the damage process acting inside the propagation cone 
is uncorrelated with the front dynamics. 
The resulting scenario is the same as the one suggested by the 
correlation-function analysis:
for ~$\eps < \eps_c(1)$ ($\eps > \eps_c(2)$) 
both $V_F $ and $V_D $ are zero (nonzero). This confirm that 
damage-spreading velocities are reliable order parameters to 
distinguish between PP and CP.  

Inside the fuzzy  region ~$[\eps_c(1),\eps_c(2)]$~ 
both zero- and nonzero-velocity regimes finely alternate without any 
apparent regularity.
We want to stress that velocity fluctuations clearly visible in the inset 
of Fig.~\ref{fig:vel} are not an artifact following from either statistical
uncertainty or finite size effect.
In fact, the initial conditions have been taken after discarding a sufficiently 
long transient, which, in some cases, amount to more than 20,000 iterations. 
In particular, the transient has been estimated from the relaxation 
properties of the ensemble average of the variables $x_i$. 
Moreover, we have increased  the system size untill we found evidence 
that the velocity does not depend of $L$. 
This means that in some ``critical'' cases we had to work with lattices 
of 6,000 sites. 
Finally, the simulations have been let
evolve for a sufficiently long time (up to $10^5$ iterations) 
to accurately determine the asymptotic behaviour.  

It is  interesting to notice that in those ambiguous cases where $C_T(\tau)$
was apparently decaying to a finite value $V_F$ and $V_D$ are strictly 
zero. This confirms the conjecture that these are truly ordered regimes.  

\section{Information-theoretic analysis}

The different features of the patterns in the fuzzy region 
indicate that the underlying dynamical mechanism is associated
to a sequence of structural changes of the phase space.
A proper method for quantifying this scenario amounts to studying the 
probability distribution function of the state variables $x_i$'s
\be
P(x) = \int_{0}^{1} dx_1 .... \int_{0}^{1} dx_L\; \rho(x_1,...x_L)
\,\delta(x - x_1) \quad ,
\label{eq:distr}
\ee
where, $\rho(x_1,...x_L)$ represents the unknown invariant measure 
generated under the evolution law (\ref{eq:cml})~.
If one assumes that $\rho$ is defined by averaging over the usual 
ensemble of initial conditions, then translation invariance is automatically 
ensured and  the choice of the non-dummy variable (here $x_1$) is irrelevant. 
The histograms of $P(x)$, shown in Fig.~7, have been obtained by averaging 
over 500 initial conditions and over the whole time span of the simulations 
(obviously, after discarding a suitable transient). In CP, where 
the damage-spreading velocities are strictly positive, $P(x)$ exhibits 
a peaked distribution superposed to a continuous component 
(see, e.g., Fig.~\ref{fig:distr}d). 
The peaks are located in correspondence of the values of some space-time
periodic orbits, which still turn out to play a role even in the
chaotic regime. 
Below $\eps_c(2)$ the continuous component is negligible and only 
the peaked structure survives (see Fig.~\ref{fig:distr}a,b and c)~. 
This occurs irrespectively of the velocity, that may be either very small
(case b) or  strictly zero (case c).  
The practical absence of a continuous componet  in the fuzzy region, even if
$V_F$ and $V_D$ are positive, leaves open the question whether this is 
due to an insufficient spatial resolution.
 
More refined information can be obtained by partitioning the unit interval 
into subintervals of equal length $\Delta$ and thereby computing the entropy 
\be
S(\Delta) = \sum_i \mu_i \ln \mu_i \quad,
\label{eq:entr}
\ee
where $\mu_i$ is the integral of $P(x)$  over the $i$-th subinterval.
The scaling behaviour of $S(\Delta)$ yields the information dimension
\be
   D_0 = \lim_{\Delta \to 0} \frac{S(\Delta)}{\ln \Delta} \quad,
\ee
which is a natural indicator quantifying the strength of the apparent
singularities. In CP, $D_0$ is steadily close to 1, 
confirming that on sufficiently fine scales the distribution is continuous
(the distance from 1 is, indeed, not appreciable). In the fuzzy transition 
region, $D_0$ exhibits irregular oscillations, while below $\eps_c(1)$
one observes a smooth tip followed by a sharp decay (see 
Fig.~\ref{fig:f-dim})~. Accordingly, in PP, $D_0$ cannot be considered as a
meaningful order parameter since the information 
dimension can be as large as in CP. Notice that 
the origin of a strictly positive dimension $D_0$, even in PP, stems 
from the irregular spatial structure irregular  alternancy of periodic 
stripes. 
In fact, when the time evolution of the CML is periodic, one can imagine 
to obtain the same invariant measure upon iterating in space  model 
(\ref{eq:cml}), after imposing periodic boundary conditions in time. 
In this perspective, one can conjecture that, at variance with the 
time evolution, the spatial iteration yields a positive Lyapunov exponent. 
Accordingly $D_0$ should be read as the fractal dimension of the corresponding 
strange repeller. A quantitative verification of these ideas, 
requiring special care in dealing with the escape rate from the
repeller,  will be performed elsewhere.

Nonetheless, we have refined our analysis by projecting the
invariant measure $\rho$ onto higher dimensional spaces, 
\be
  P(x_1,\ldots,x_E) = \int_0^1 dx_{E+1} \ldots \int_0^1 dx_L
   \rho(x_1,\ldots,x_L)   \quad, 
\ee
and studying the corresponding fractal dimensions. 
Notice that for $E=1$, the above equation reduces to Eq.~(\ref{eq:distr}).
An effective study of $P(x_1,\ldots,x_E)$ can be performed by interpreting
a spatial configuration as a time-series and thereby applying embedding
techniques (see, e.g., \cite{Gr}). We have applied the Grassberger-Procaccia 
method \cite{GraPro} to configurations of length $M=10^5$. 
The technique consists in calculating the correlation integral 
\be
{\cal N}(E,\Delta) = \frac{2}{M(M-1)} \sum_{i < j} 
\Theta(\Delta - \|x_i - x_j\|_E) 
\label{eq:int_corr}
\ee
where $\Theta(t)$ is the Heaviside function and $\|...\|_E$ is some 
norm in an $E$-dimensional space. The estimation of the correlation 
dimension $D_2$ can be obtained from the asymptotic behavior of the
effective dimension
\be
 D_2(E,\Delta) =  \frac{\de{\cal N}}{\de \ln\Delta} 
\ee
as $\Delta\to 0$. The results for different values of $\eps$ and embedding 
dimension $E=1,2,3$ are reported in Fig.~\ref{fig:embedd}. At variance with 
the previous cases, this analysis seems to identify only one transition 
point, at $\eps_d = 0.301$  which is definitely larger than $\eps_c(2)$ 
(the largest $\eps$-value at which the damage spreading velocity drops to zero).
In fact, above $\eps_d$, $D_2$ increases with the embedding dimension
(see, e.g., Fig.~\ref{fig:embedd}e, f), while, below $\eps_d$, the fractal 
dimension is independent of $E$ (see Fig.~\ref{fig:embedd}a-d). 
This means that, in PP, the invariant measure has a finite fractal 
dimension, which can be determined already for $E=1$. 
This is not so surprising as long as the temporal evolution is 
periodic; it is less obvious that in the aperiodic regimes occurring inside
the fuzzy region one has the same behaviour. Another question that is left
open by the simulations is the actual dimension of the invariant measure 
deeply inside CP: it is not clear whether the dimension is
finite although very large, or if it is infinite as in standard space-time
chaos. Unfortunately, at present, there are no theoretical arguments that
can help clarifying this question: in particular, we cannot resort to the 
Kaplan-Yorke conjecture for concluding that the dimension should be 
proportional to the system-size and thus infinite in the thermodynamic limit.

\section{Conclusions}

A peculiar non-equilibrium transition between an ordered and a disordered 
dynamical phase has been identified in a one dimensional Lyapunov-stable CML 
model. Various observables have been considered as possible order 
parameters. In fact, either the exponential decay rate of the 
temporal correlation functions ($\theta$) in the chaotic phase, 
or the damage spreading 
velocities ($V_F$, $V_D$) and the fractal dimension ($D_0$) agree in 
identifying a phase transition region extending over a finite, even if narrow, 
interval in the control parameter space. The existence of such a ``fuzzy'' 
region represents one of the main interesting aspects of the model at hand, 
in which transient ``chaotic'' evolution has been already detected \cite{PLOK}. 
Analogies and similiraties with other kinds of complex models have been 
discussed throughout the paper, leading to the conclusion that this phase 
transition is characterized by specific features, that make it different from 
any other known scenario.

Let us finally stress that, besides the above mentioned order parameters, there
exists at least one observable that, seemingly, identifies a true 
phase transition point. This was found by studying the fractal dimension of
spatially embedded configurations. The motivation for such a transition
and its connection with the dynamics occurring in the fuzzy region identified
by the other order parameters is not understood. We believe that the absence 
of sources of randomness (either stochastic or chaotic) is at the
origin of this complex scenario.  We expect that clarifying this question
could shed some light on the very nature of complex behaviour at the edge of 
chaos.

\acknowledgments
We are indebted with Y. Cuche for his contribution to the early stages 
of this work. We want to acknowledge useful discussion with G. Grinstein and
R. Kapral. We also thank I.S.I. in Torino for the kind hospitality during 
the workshop of the EU HC\&M Network ERB-CHRX-CT940546 on ``Complexity and
chaos'', where part of this work was performed.
    

\begin{figure}
\caption{Four space-time patterns generated by the coupled-map lattice
(1) for different values of the coupling constant $\eps$.
In all cases, time flows downwards and the patterns, 
200 $\times$ 300 wide, are extracted from the evolution of a 
3000-site lattice with the same randomly chosen initial condition. 
Case a) displays an ordered regime for  $\eps=0.2998$; the
more complex pattern in b) is obtained for $\eps=0.3004$; c) displays 
quasi ordered pattern generated for $\eps=0.3005$;
a totally disordered regime is shown in d) for $\eps=0.304$~.}
\label{fig:patt}
\end{figure}

\begin{figure}
\caption{Average transient time $\langle T_r(L) \rangle$ (a) and average 
period $\langle T_p(L) \rangle$ (b) versus $\eps$. Both averages are performed 
over 200 random initial conditions. Solid and dashed curves refer to a chain 
length $L=50,\;40$, respectively.}
\label{fig:trans}
\end{figure}

\begin{figure}
\caption{Spatial correlation function (5) for different values
of $\eps$: $0.2998$ (full line), $0.3005$ (dashed line), $0.3040$ 
(dot-dashed line). All the curves are obtained by averaging over $500$ 
random initial conditions and $10^5$ time steps.}  
\label{fig:corrx}
\end{figure}
 
\begin{figure}
\caption{Temporal-correlation functions for the same values of $\eps$ 
considered in Fig.~1. For clarity reasons, in a) and c), the 
points are not connected by lines. In the ordered regime (a) the correlation 
function does not decay and exhibits essentially period-2 oscillations;
in the complex pattern (b), a slow decay to zero is observed;  
in the ordered regime (c), inside the critical region, period-2 oscillations
co-exist with a temporal decay to a finite asymptotic value;
in the chaotic region (d), period-two oscillations modulate a much faster 
decay than in case (b).}
\label{fig:corrt}
\end{figure}

\begin{figure}
\caption{Damage spreading patterns corresponding to the same
$\eps$-values considered in Fig.~1.}
\label{fig:def-patt}
\end{figure}

\begin{figure}
\caption{The damage spreading velocities $V_F$ and $V_D$ vs. $\eps$. 
The inset amplifies the critical region $[\eps_c(1),\eps_c(2)]$,
where periodic and chaotic regimes irregularly alternate.}
\label{fig:vel}
\end{figure}

\begin{figure}
\caption{Probability density $P(x)$ of the site-variable  for 
the same $\eps$-values considered in Fig.~1.} 
\label{fig:distr}
\end{figure}

\begin{figure}
\caption{Information dimension $D_0$ of the single variable invariant
measure vs. $\eps$.}
\label{fig:f-dim}
\end{figure}

\begin{figure}
\caption{Effective dimension $D_2$ vs. $\ln \Delta$ for 
embedding dimension $E=1,2,3$ (solid, dashed and dot-dashed lines, respectively)
for different values of $\eps$: $0.2998$ (a), $0.3004$ (b), $0.3005$ (c), 
$0.3015$ (d), $0.302$ (e), and $0.304$ (f).}     
\label{fig:embedd}
\end{figure}


\begin{thebibliography}{99}
\bibitem{comm} Evidently, the term ``ordered'' refers to the
temporal behaviour of the entire lattice in a frame moving with the velocity
of the surviving DWs. Such a velocity can be different from zero even
in a dynamical system with left-right symmetry, if the chosen solution
breaks this symmetry  yielding tilted stripes (see also \cite{PB}).
\end{thebibliography}
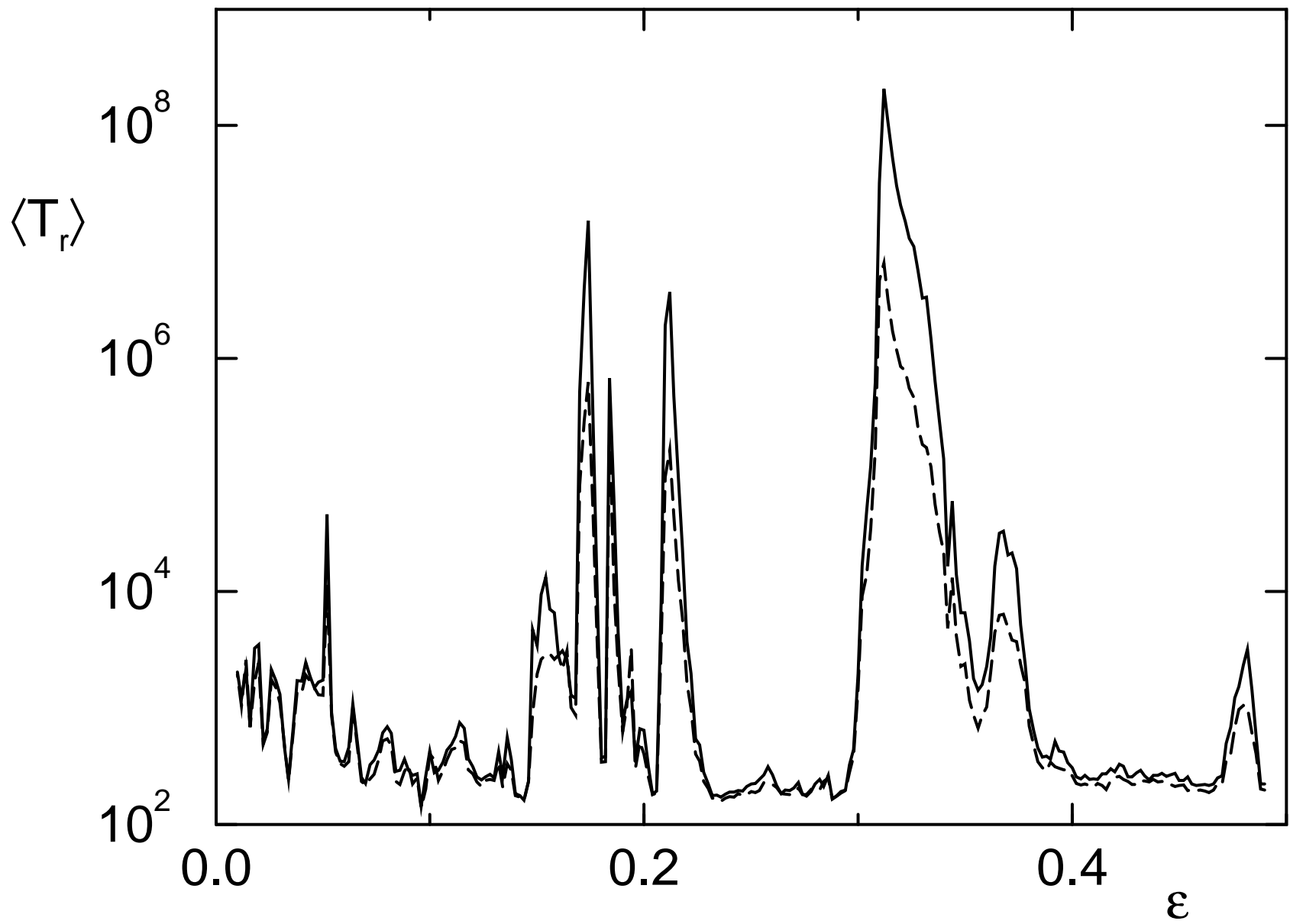
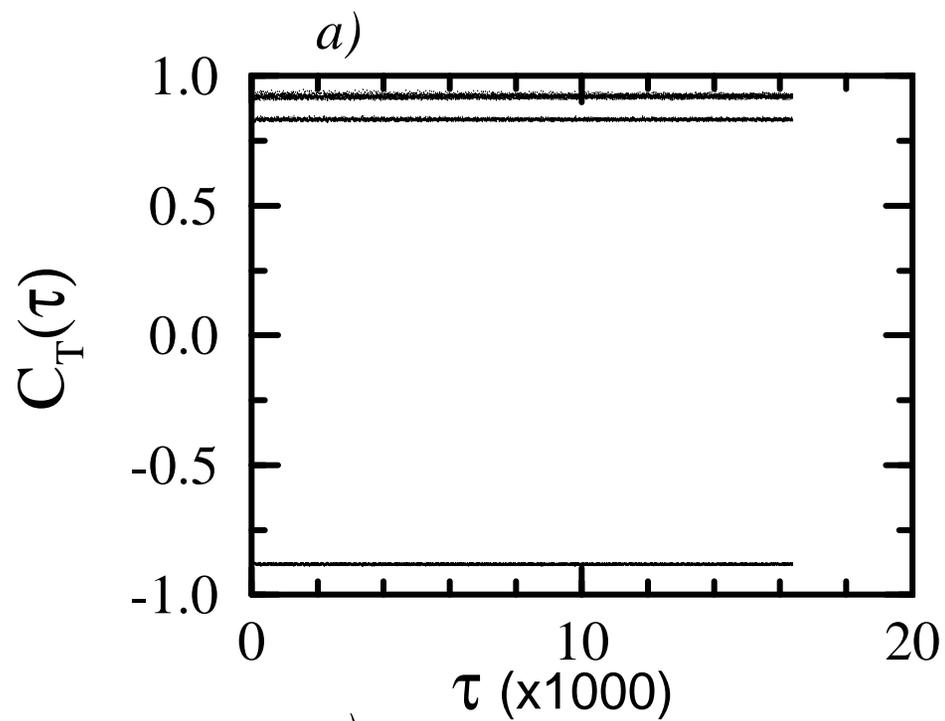
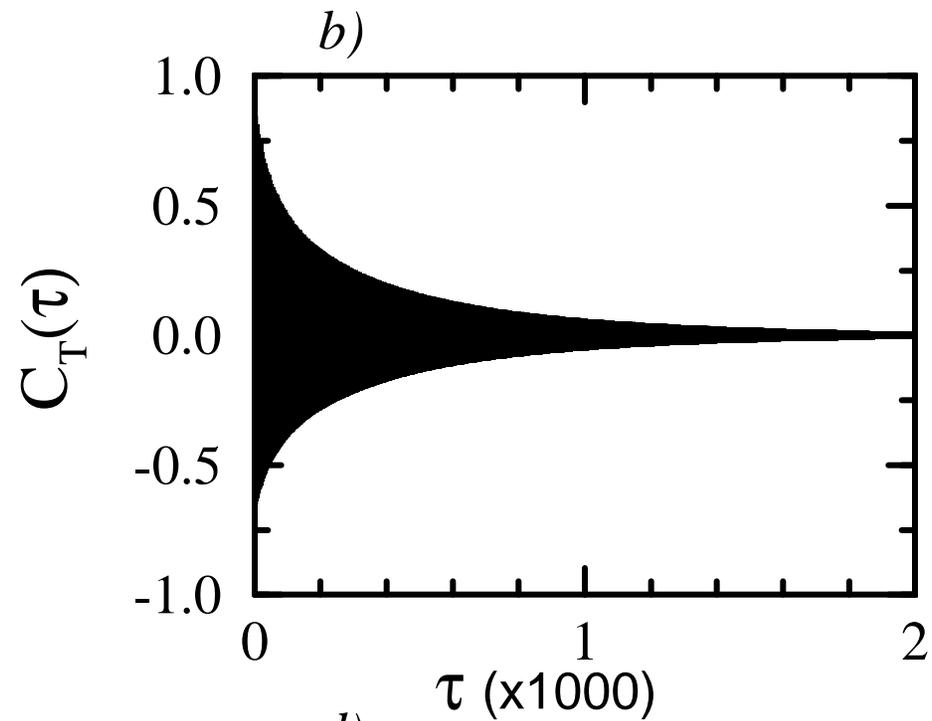
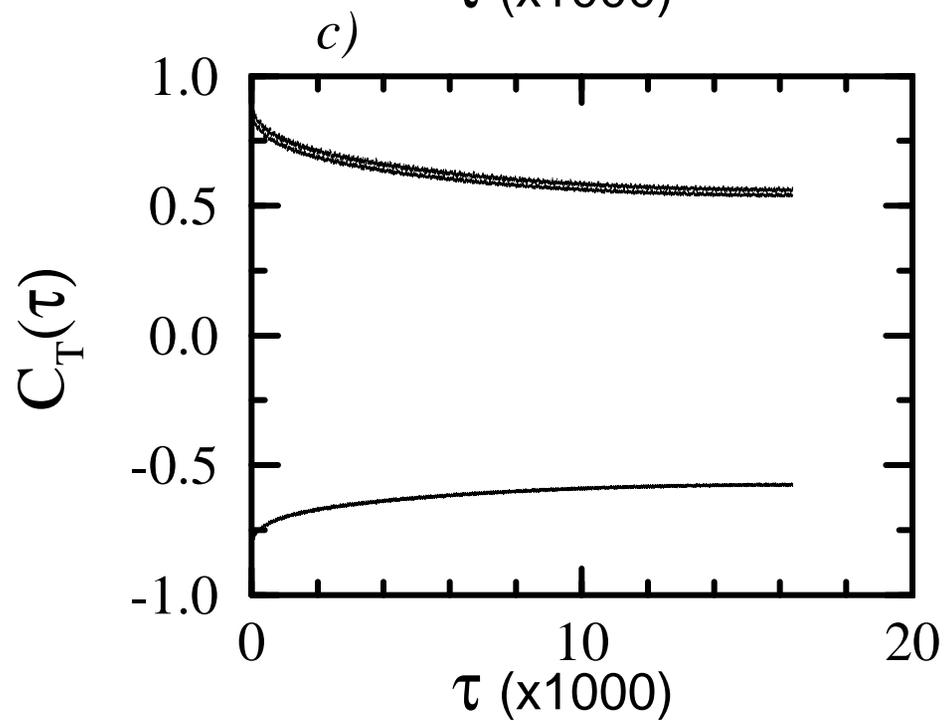
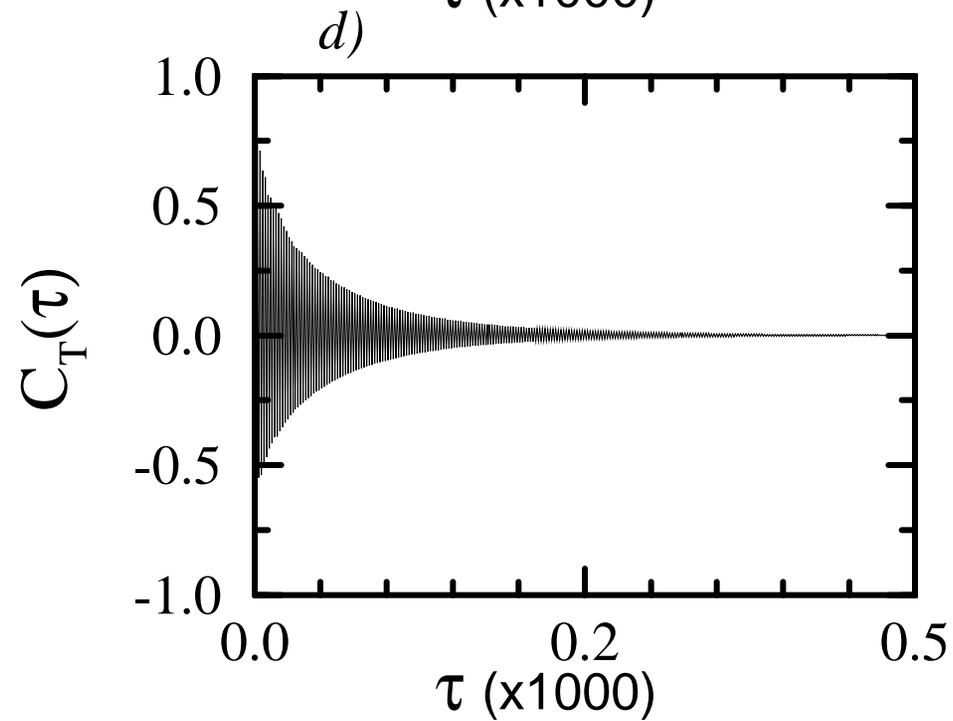
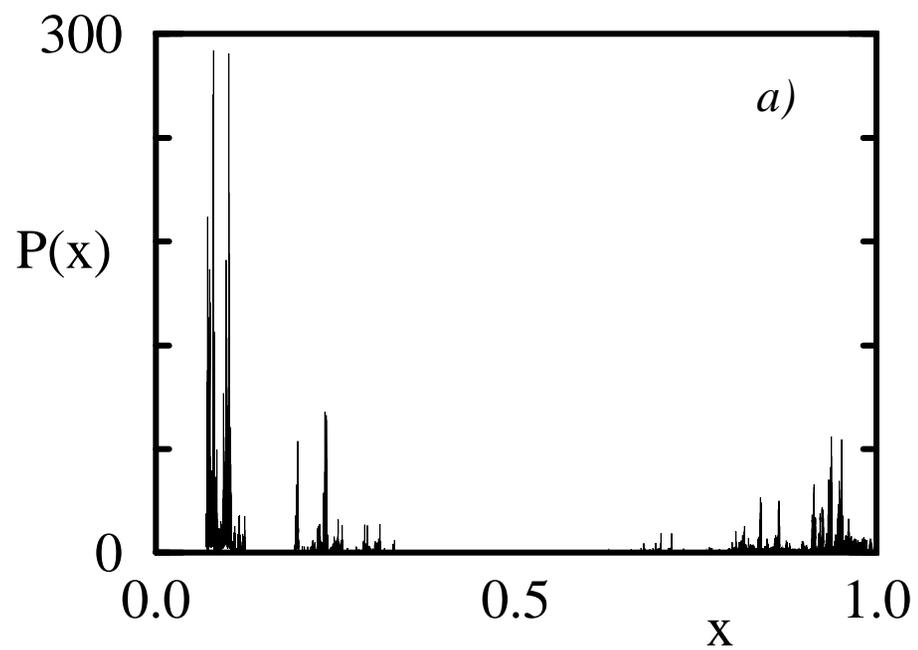
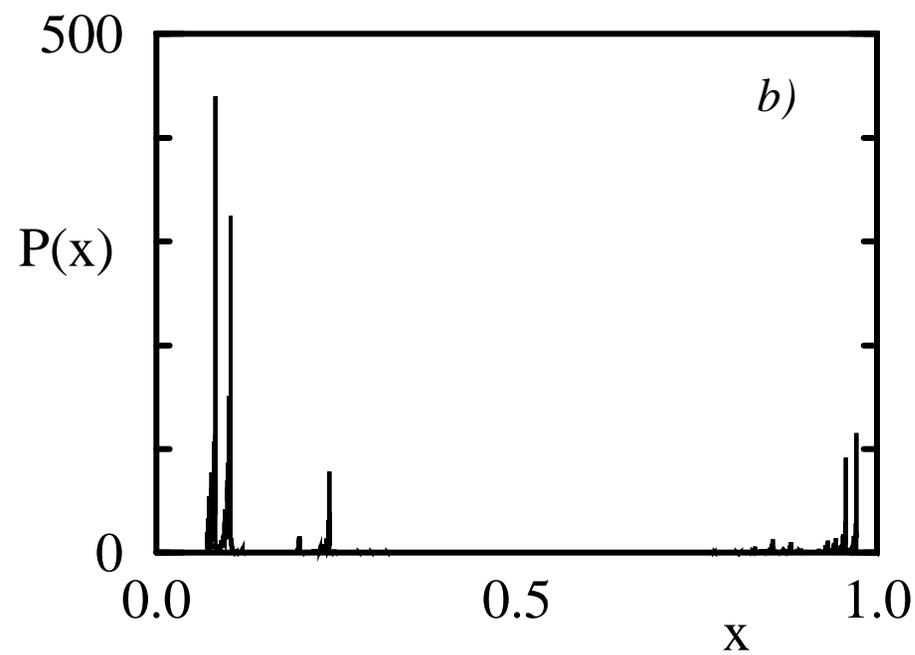
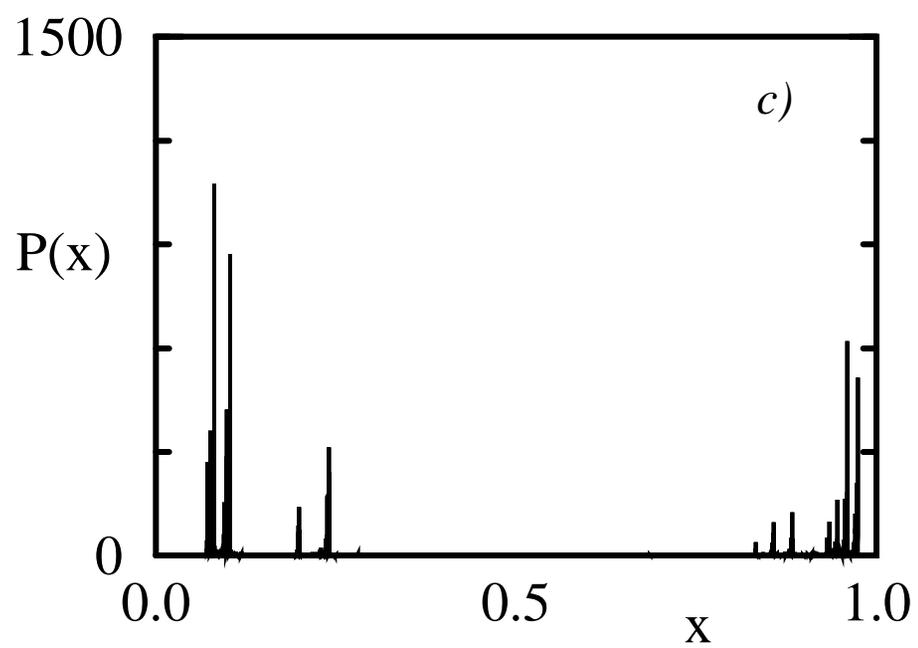
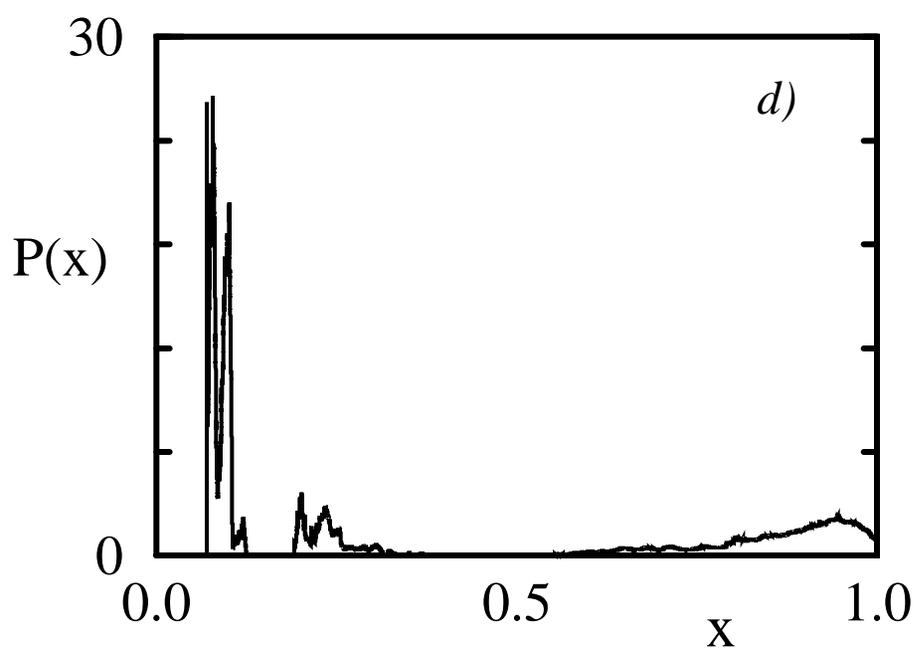
\end{document}